\documentclass[shortnote,twocolumn]{jpsj3}
\usepackage{amsmath}
\usepackage{newtxtext}
\usepackage[varg]{newtxmath}
\usepackage{bm}
\usepackage{graphicx}
\usepackage{dcolumn}
\usepackage{color}

\title{Electric-Field-Induced Antiferromagnetic Insulating State in a Metallic Ferromagnet}
\author{Atsushi Ono\thanks{ono@cmpt.phys.tohoku.ac.jp} and Sumio Ishihara}
\inst{Department of Physics, Tohoku University, Sendai 980-8578, Japan} 

\abst{
We show that a static electric field induces the transition from a ferromagnetic metal to an antiferromagnetic insulator owing to the Bloch oscillation of conduction electrons.
In the steady state, the electric current is inversely proportional to the applied electric field, implying the nonperturbative insulating nature that is different from the Wannier--Stark localization.
Possible experimental realization based on recent terahertz pulse sources is discussed.
}

\begin{document}
\maketitle

Ultrafast manipulation of magnetic order has been of importance in the field of spintronics.
Thus far, most of them are based on the heating effect or the inverse magneto-optical effects by optical pulse irradiation.~\cite{Kirilyuk2010,Walowski2016}
Another strategy is to use the nonthermal effects of light, e.g., the photoinduced magnetic phase transition~\cite{Ishihara2019} and the Floquet engineering.~\cite{Mentink2017,Oka2018}
Recently, the authors reported that the ferromagnetic (FM) double-exchange interaction turns into the antiferromagnetic (AFM) interaction under strong continuous wave (cw).~\cite{Ono2017,Ono2018,Ono2019}
This is in contrast to the photoinduced AFM-insulator--to--FM-metal transition attributed to the photocarrier-mediated FM interaction.~\cite{Ishihara2019}
Here, we show that a static electric field can also induce the transition to the AFM steady state in which the electric current does not flow.
This provides a more feasible way by means of terahertz light sources because the transition timescale and the required electric-field magnitude are within reach of state-of-the-art experiments.

We consider the double-exchange model on the square lattice.
The Hamiltonian is given by
\begin{align}
\mathcal{H} = -h \sum_{\langle ij \rangle s} c_{is}^\dagger c_{js} - J \sum_{iss'} \bm{S}_i \cdot \bm{\sigma}_{ss'} c_{is}^\dagger c_{is'},
\label{eq:hamiltonian}
\end{align}
where $c_{is}^\dagger \ (c_{is})$ is the creation (annihilation) operator of a conduction electron with spin $s \ (={\uparrow},{\downarrow})$ at site $i$, and $\{\sigma^x,\sigma^y,\sigma^z\}$ are the Pauli matrices.
The localized spins are treated as classical unit vectors ($\vert \bm{S}_i\vert = 1$).
The first term describes the electron hopping, and the second term represents the on-site Hund coupling.
We adopt a finite-size cluster with the total number of sites $N=8\times 8$, that of electrons $N_\mathrm{e}$, and the electron density $n_\mathrm{e}=N_\mathrm{e}/N$.
We introduce a static electric field as a simplification of few-cycle terahertz pulses, focusing on a timescale up to $1~\mathrm{ps}$.
In the temporal gauge, the static electric field $\bm{F}_0$ is given by $\bm{F}_0=-\partial_t \bm{A}(t)$, where $\bm{A}(t)=-\bm{F}_0t$ $(t\geq 0)$ is the vector potential.
This is different from cw or impulsive excitations that were discussed in Ref.~\citen{Ono2017}.
The vector potential is incorporated in the Peierls phase as $h c_{is}^\dagger c_{js} \mapsto h \exp[\mathrm{i} \bm{A}(t)(\bm{r}_i-\bm{r}_j)] c_{is}^\dagger c_{js}$ with $\bm{r}_i$ being the position of the $i$th site.
The electric field is applied in the diagonal direction of the square lattice, i.e., $\bm{F}_0=(F_0,F_0)$.
Throughout this paper, the initial state at $t=0$ is taken to be a FM-metallic ground state with $J=5h$ and $n_\mathrm{e}=0.5$;
the FM-to-AFM transition shown later occurs in a wide range of $J$ and $n_{\mathrm{e}}$.
The nearest-neighbor hopping amplitude $h$, the reduced Planck constant $\hbar$, the electron charge $e$, and the lattice constant $a$ are set to unity.
The units of time ($\hbar/h$), the electric field ($h/ea$), and the electric current density ($eh/\hbar a^2$), respectively correspond to $1.32~\mathrm{fs}$, $10~\mathrm{MV/cm}$, and $4.9\times 10^{10}~\mathrm{A/cm^2}$, for $h=0.5~\mathrm{eV}$ and $a=0.5~\mathrm{nm}$.
We perform numerically exact calculations of the real-time evolution of the electrons and spins in the finite cluster using the method presented in Refs.~\citen{Chern2018} and \citen{Luo2019}.
The density matrix of the electrons, $\rho_{is,js'}=\langle c_{js'}^\dagger c_{is} \rangle$, follows the von~Neumann equation, while the time evolution of the localized spins $\{\bm{S}_i\}$ is governed by the Landau--Lifshitz--Gilbert equation, $\partial_t \bm{S}_i = \bm{S}_i \times \bm{b}_{i} - \alpha \bm{S}_i \times \partial_t\bm{S}_i$, where $\bm{b}_i=-\partial\langle\mathcal{H}\rangle/\partial\bm{S}_i$ is the effective field and $\alpha$ is the damping constant.
These equations are simultaneously solved by using the fourth-order Runge--Kutta method with a time step $\delta t = 0.001$.
In the initial state, small randomness is introduced in the localized spins so that the spin structure can be changed by the external field via the conduction electrons; the maximum deviation of the polar angle is $\delta\theta = 0.1$ rad.~\cite{Ono2017}
The periodic (antiperiodic) condition is imposed along the $x$ ($y$) direction, which stabilizes the FM initial state.

\begin{figure}[t]
\centering
\includegraphics[width=0.94\columnwidth]{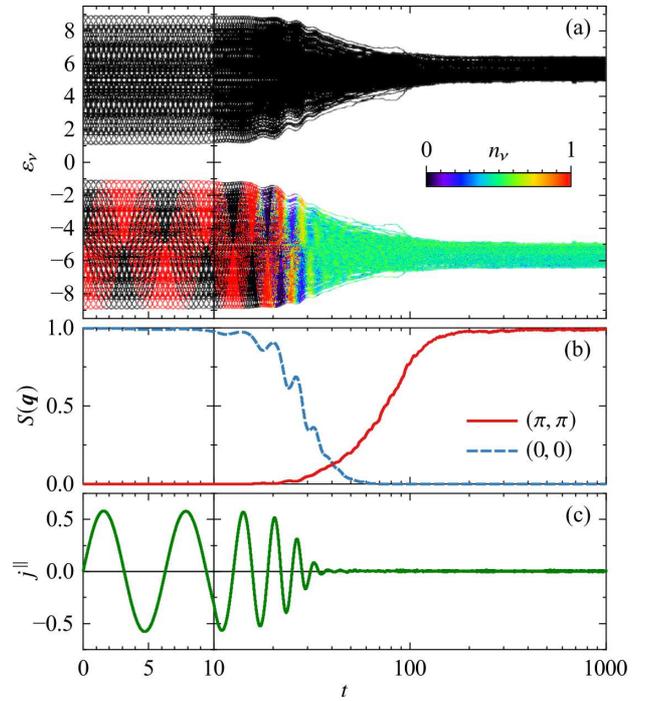}
\caption{
Time profiles of (a)~the single particle energy $\varepsilon_\nu$ and its occupation number $n_\nu$, (b)~the structure factor $S(\bm{q})$, and (c)~the electric current density $j^\parallel$.
The parameter values are $F_0=1$ and $\alpha=1$. \vspace{-6mm}
}
\label{fig:timeprofile}
\end{figure}

In Fig.~\ref{fig:timeprofile}, we show the time profile of the single-particle energy $\varepsilon_\nu$ and its occupation number $n_\nu$, the spin structure factor $S(\bm{q})=N^{-2}\sum_{ij}\bm{S}_i{\cdot}\bm{S}_j \exp[\mathrm{i}\bm{q}(\bm{r}_i-\bm{r}_j)]$, and the electric current density $j^\parallel = -N^{-1}\partial \langle \mathcal{H} \rangle/\partial A^\parallel$ that is parallel to $\bm{F}_0$.
At early times ($t\lesssim 10$), $j^\parallel$ shows the Bloch oscillation with a time period of $T = 2\pi \hbar/(eF_0a) = 6.28$.
Note that this oscillation is not induced by cw.~\cite{Ono2017}
The electron bandwidth and $S(0,0)$ decrease most quickly when the electron population is inverted in the lower band;
they slightly return toward the initial values when the population is normally distributed.
This is seen in Figs.~\ref{fig:timeprofile}(a) and \ref{fig:timeprofile}(b) at intermediate times ($10\lesssim t \lesssim 30$).
After $S(0,0)$ and $j^\parallel$ fade away at $t\sim 30$, the electron distribution becomes almost uniform in the lower band and $S(\pi,\pi)$ gradually develops.
The steady state is realized at late times ($t\gtrsim 200$), where $S(\pi,\pi) \approx 1$ and $j^\parallel \approx 0$, indicating the long-range (up to the cluster size at least) AFM correlation and the insulating nature.
This insulating behavior at the late times is different from the Wannier--Stark localization since the Bloch oscillation is not observed any longer.

\begin{figure}[t]
\centering
\includegraphics[width=0.95\columnwidth]{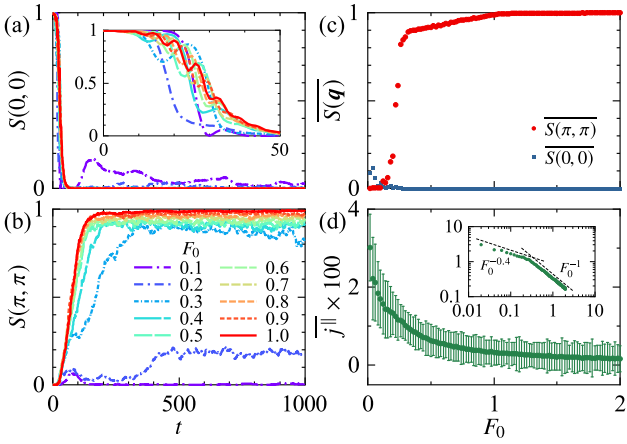}
\caption{
(a, b)~Time profiles of the structure factor.
The inset of (a) is the enlarged view from $t=0$ to $50$.
(c)~Time-averaged values of the structure factors and (d)~those of the electric current density in the steady states ($t=600$--$1000$).
The errorbars in (d) represent the standard deviations of the time averages.
The inset of (d) is a logarithmic plot of $\overline{j^\parallel}$.
The Gilbert damping constant is $\alpha=1$ in (a)--(d). \vspace{-6mm}
}
\label{fig:timeaverage}
\end{figure}

The FM and AFM structure factors, $S(0,0)$ and $S(\pi,\pi)$, are shown in Figs.~\ref{fig:timeaverage}(a) and \ref{fig:timeaverage}(b), respectively, for different values of $F_0$.
Around $t=20$--$30$, $S(0,0)$ vanishes and $S(\pi,\pi)$ starts to increase.
While the asymptotic values of $S(0,0)$ at the late times are less than $0.05$ irrespective of $F_0=0.1$--$1$, that of $S(\pi,\pi)$ monotonically increases with $F_0$.
Figures~\ref{fig:timeaverage}(c) and \ref{fig:timeaverage}(d) show the time averages of the structure factor and the electric current, $\overline{S(\bm{q})}$ and $\overline{j^\parallel}$, respectively, in the steady states from $t=600$ to $1000$.
The AFM steady states are stabilized when $F_0 \gtrsim 0.2$.
Even for $F_0\lesssim 0.2$, $\overline{S(0,0)}$ is smaller than $0.1$, which indicates the melting of the initial FM order.
The time-averaged current in Fig.~\ref{fig:timeaverage}(d) shows the power-law behavior in $F_0$, i.e., $\overline{j^\parallel} \propto F_0^{-0.4}$ for $F_0\lesssim 0.2$ and $\overline{j^\parallel} \propto F_0^{-1}$ for $F_0\gtrsim 0.2$.
This means that the differential conductivity $\partial j^\parallel/\partial F_0$ is negative and the change from the FM metal to the AFM insulator is a nonperturbative effect because of the singularity at $F_0=0$.

\begin{figure}[t]
\centering
\includegraphics[width=0.95\columnwidth]{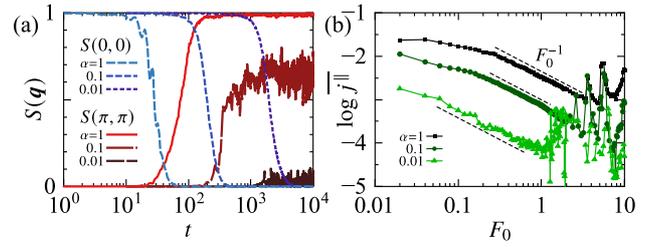}
\caption{
(a)~Time profiles of the structure factor for $\alpha=0.01,0.1,1$.
The field magnitude is $F_0=1$.
(b)~The time-averaged electric current density in the steady states ($t=8000$--$10000$). \vspace{-6mm}
}
\label{fig:alpha}
\end{figure}

The important phenomenological parameter that governs the timescale of the transition is the damping constant $\alpha$, whose realistic value is $\alpha\sim 0.01$.~\cite{Luo2017,Ghosh2019}
Figure~\ref{fig:alpha}(a) shows the time profile of $S(\bm{q})$ for $\alpha=0.01,0.1,1$ and $F_0=1$.
The time at which the FM order begins to melt is almost linear in $\alpha^{-1}$, which is $t \lesssim 10^4\sim 10~\mathrm{ps}$ when $\alpha=0.01$.
However, as shown in Fig.~\ref{fig:alpha}(a), both the FM and AFM structure factors are below $0.1$ for $\alpha=0.01$ in the steady state, which implies the paramagnetic structure.
The further calculations for $F_0>2$ reveal that $\overline{S(\pi,\pi)}$ shows nonmonotonic dependence on $F_0$ and takes values greater than $0.9$ at some magnitudes (not shown).
This might be attributed to the treatment of the relaxation in the present calculations, where it is taken into account only through the Gilbert damping.
The total energy takes its minimum when the magnetic structure is AFM and the electrons uniformly occupy the lower band.~\cite{Ono2017}
Therefore, we conjecture that the AFM state may be stabilized in a broader range in $F_0$ if other relaxation processes (e.g., scattering by the electron-electron interaction) are considered.
Figure~\ref{fig:alpha}(b) displays the time-averaged value of $j^{\parallel}$, which is proportional to $F_0^{-1}$ even for $\alpha=0.01$.
Thus, the insulating behavior is observed regardless of the appearance of the AFM structure.

The AFM transition shown above can be verified by means of intense terahertz pulse sources.
A candidate material is the perovskite manganite, $\textrm{La}_{1-x}\textrm{Sr}_{x}\textrm{MnO}_{3}$, where the FM-metallic phase is realized by the double-exchange interaction.
A cycle of terahertz pulses is of the order of $1~\mathrm{ps}$ and the peak intensity has reached $100~\mathrm{MV/cm}$,~\cite{Koulouklidis2020} which justifies the treatment of the pulse as the static field in the timescale concerned in this study.
Although the power-law behaviors shown in Figs.~\ref{fig:timeaverage}(d) and \ref{fig:alpha}(b) indicate that the FM-metallic state is unstable against infinitesimal $F_0$, this is an artifact of the present calculations in which the intraband relaxation time $\tau$ is infinite as long as the system is fully FM.
Hence, in real materials, the Bloch-oscillation period $T\propto F_0^{-1}$ needs to be shorter than the relaxation time $\tau$ to avoid the thermalization of the electron distribution.
The estimate of this threshold is at most $F_0 \sim 2\pi\hbar/(ea\tau) \approx 80~\mathrm{MV/cm}$ for $\tau=1~\mathrm{fs}$ and $a=0.5~\mathrm{nm}$.

In summary, we have shown the static-electric-field-induced transition from a FM metal to an AFM insulator.
The instability of the FM-metallic state is due to the Bloch oscillation of the conduction electrons, unlike the nonthermal distribution in cw-induced Floquet states.~\cite{Ono2017,Ono2018}
The electric current is inversely proportional to the electric field in the steady states, which implies the insulating nature that is different from the Wannier--Stark localization.
This transition can be achieved by the intense terahertz pulse sources.

\begin{acknowledgment}
This work was supported by JSPS KAKENHI Grant Nos. JP19K23419, JP20K14394, JP17H02916, JP18H05208, and JP20H00121.
Some of the numerical calculations were performed using the facilities of the Supercomputer Center, the Institute for Solid State Physics, the University of Tokyo.
\end{acknowledgment}
\vspace{-7mm}

\bibliography{reference}

\end{document}